\documentstyle[12pt]{article}
\catcode`\@=11

\def\@normalsize{\@setsize\normalsize{15pt}\xiipt\@xiipt
\abovedisplayskip 14pt plus3pt minus3pt%
\belowdisplayskip \abovedisplayskip
\abovedisplayshortskip  \z@ plus3pt%
\belowdisplayshortskip  7pt plus3.5pt minus0pt}

\def\small{\@setsize\small{13.6pt}\xipt\@xipt
\abovedisplayskip 16pt plus3pt minus3pt%
\belowdisplayskip \abovedisplayskip
\abovedisplayshortskip  \z@ plus3pt%
\belowdisplayshortskip  7pt plus3.5pt minus0pt
\def\@listi{\parsep 4.5pt plus 2pt minus 1pt
            \itemsep \parsep
            \topsep 9pt plus 3pt minus 3pt}}

\@twosidetrue

\catcode`@=12

\catcode`\@=11

\def\thesection{\Roman{section}}

\def\ps@headings{\def\@oddfoot{}\def\@evenfoot{}}
\def\@oddhead{\hbox{}\hfill
        \makebox[.5\textwidth]{\raggedright\ignorespaces --\thepage{}--
        \hfill }}
\def\@oddhead{\hbox{}\hfill --\thepage{}-- \hfill}
\def\@evenhead{\@oddhead}

\ps@headings
\catcode`\@=12

\relax
\newcounter{appendix}
\def\appendix{\par
 \reset{equation}
 \addtocounter{appendix}{1}
 \def\thesection{\Alph{appendix}.}
 \def\ksection{\Alph{appendix}}}

\begin{document}
\newcommand{\dal}{\raisebox{0.085cm}
{\fbox{\rule{0cm}{0.07cm}\,}}}
\newcommand{\gef}{G_{\rm eff}}
\newcommand{\lef}{\Lambda_{\rm eff}}
\newcommand{\al}{\alpha^{\prime}}
\newcommand{\mst}{M_{\scriptscriptstyle \!S}}
\newcommand{\mpl}{M_{\scriptscriptstyle \!P}}
\newcommand{\alg}{\alpha_{\scriptscriptstyle G}}
\newcommand{\bg}{\beta_{\scriptscriptstyle G}}
\newcommand{\bl}{\beta_{\scriptscriptstyle \Lambda}}
\newcommand{\half}{\textstyle\frac{1}{2}}
\newcommand{\trz}{\textstyle\frac{1}{3}}
\newcommand{\shalf}{{\textstyle\frac{1}{\sqrt{2}}}}
\newcommand{\frr}{{\textstyle\frac{1}{4}}}
\newcommand{\czt}{{\textstyle\frac{1}{4}}}
\newcommand{\sze}{{\textstyle\frac{1}{6}}}
\newcommand{\err}{{\textstyle\frac{1}{8}}}
\newcommand{\osi}{{\textstyle\frac{1}{8}}}
\newcommand{\srr}{{\textstyle\frac{1}{16}}}
\newcommand{\gn}{\frac{1}{16\pi G}}
\newcommand{\gr}{G_{\!\rm eff}}
\newcommand{\eff}{\Gamma_{\!\rm eff}}
\newcommand{\lr}{\Lambda_{\rm eff}}
\newcommand{\gnn}{\frac{1}{32\pi G}}
\newcommand{\dv}{\int{\rm d}^4x\sqrt{g}}
\newcommand{\lac}{\lambda_{\scriptscriptstyle G}}
\newcommand{\act}{\widetilde{\Gamma}}
\newcommand{\lv}{\left\langle}
\newcommand{\rv}{\right\rangle}
\newcommand{\ph}{\varphi}
\newcommand{\sbar}{\,\overline{\! S}}
\newcommand{\tbar}{\overline{T}}
\newcommand{\ybar}{\overline{Y}}
\newcommand{\nbar}{\,\overline{\! N}}
\newcommand{\qbar}{\,\overline{\! Q}}
\newcommand{\phb}{\overline{\varphi}}
\newcommand{\cm}{Commun.\ Math.\ Phys.~}
\newcommand{\pr}{Phys.\ Rev.\ D~}
\newcommand{\pl}{Phys.\ Lett.\ B~}
\newcommand{\np}{Nucl.\ Phys.\ B~}
\newcommand{\e}{{\rm e}}
\newcommand{\suf}{\sum_{i,j=1}^{M}}
\newcommand{\gsi}{\,\raisebox{-0.13cm}{$\stackrel{\textstyle
>}{\textstyle\sim}$}\,}
\newcommand{\lsi}{\,\raisebox{-0.13cm}{$\stackrel{\textstyle
<}{\textstyle\sim}$}\,}
\newcommand{\lag}{{\cal L}^{\rm eff}}
\newcommand{\hh}{^{(h)}}
\newcommand{\ab}{^{(a)}}
\newcommand{\ren}{_{\scriptscriptstyle R}}
\newcommand{\res}{\langle{\rm Re}S\rangle}
\newcommand{\sre}{S_{\!\scriptscriptstyle R}}
\newcommand{\tre}{T_{\!\scriptscriptstyle R}}
\newcommand{\apr}{\widetilde{A}}
\begin{titlepage}
\vspace*{-2cm}
\begin{flushright}
CERN--TH.6259/91\\[-2mm]
\end{flushright}
\vskip0.7in
\begin{center}
\large\sc
Duality Invariant Effective String Actions and
\end{center}
\begin{center}
\large\sc
Minimal Superstring Unification
\end{center}
\par \vskip 0.3in\noindent
\begin{center}
{\large D. L\"ust}\raisebox{1mm}{$^{\star}$}\\[-1mm]
{\sl CERN, CH--1211 Geneva 23, Switzerland}\\[3mm]
\end{center}
\par \vskip .60in
\begin{center}
{Abstract}\\
\end{center}
\begin{quote}
Some results are presented concerning duality invariant
effective string actions and the construction of automorphic
functions for general (2,2) string compactifications.
These considerations are applied in order to discuss the {\it minimal}
unification of gauge
coupling constants in orbifold compactifications with special emphasis
on string threshold corrections.
\end{quote}
\vskip0.9in
\begin{center}
{\it Talk presented at the Joint International Lepton-Photon and
EPS-Conference,}
\end{center}
\begin{center}
{\it Geneva, 25 July -- 1 August 1991}
\end{center}
\begin{flushleft}
\rule{5.1 in}{.007 in}\\[-3mm]
$^{\star}${\small Heisenberg Fellow}\\[2mm]
CERN--TH.6259/91\\[-2mm]
September 1991
\end{flushleft}
\end{titlepage}

Target space duality symmetries (such as $R\rightarrow\tilde R=
1/R$ \cite{KIK})
provide very strong constraints
when constructing the four-dimensional string
effective supergravity action.
As first shown in ref.
\cite{FLST} for orbifold compactifications,
the requirement of
target space modular invariance
establishes a connection between the $N=1$ effective
supergravity action
and the theory of modular functions.

Consider the one-loop string threshold corrections \cite{KAPLU}\
to the gauge
coupling constants  for heterotic string compactifications on
six-dimensional Calabi--Yau spaces \cite{CHSW}.
Massless degrees of freedom
are given by gauge singlet
moduli fields $T_i$ ($i=1,\dots ,n$).
In addition there are chiral matter
fields $\phi_i^{R_a}$ ($i=1,\dots ,h_{R_a}$)
in the ${\underline{R}}_a$ representation of the
gauge group $G=\prod G_a$.
(For (2,2) compactifications $G=E_6\times E_8$,
${\underline{R}}={\underline{27}}$ and $h_{27}=n$.)
The relevant part of the tree-level
supergravity Lagrangian is specified by
the following K\"ahler potential at lowest order in $\phi_i^{R}$:
$K=K_0(T_i,\bar T_i)+K_{ij}^{R}(T_i,\bar T_i)
\phi_i^{R}\bar\phi_j^{R}$.
As discussed in refs. \cite{DFKZ,LOUIS,CAOV}, at the one-loop level
$\sigma$-model anomalies
lead, via supersymmetry, to the following one-loop modification
of the gauge coupling constants:
\begin{equation}
{1\over g^2_a}={k_a\over g_{\rm st}^2}-{1\over 16\pi^2}
\biggl( \lbrack C(G_a)-h_{R_a}T(R_a)\rbrack K_0+
2T(R_a)\log\det K_{ij}^{R_a}\biggr) .
\end{equation}
Here $g_{\rm st}$ is the string coupling
constant and $k_a$ is the level of the corresponding Kac--Moody
algebra; $C(G_a)$ and $T(R_a)$ are the quadratic Casimirs of the
adjoint and ${\underline{R}}_a$ representation of $G_a$, respectively.

Now consider
target space duality transformations. These are discrete
reparametrizations of the moduli, $\Gamma$: $T_i\rightarrow
\tilde T_i(T_i)$, which leave the string theory invariant.
Duality transformations act as
K\"ahler transformations on the moduli K\"ah\-ler potential,
i.e.
$K_0\rightarrow K_0+\log |g(T_i)|^2$.
In addition,
$\Gamma$ acts also non-trivially on the K\"ahler metric $K_{ij}^{R}$.
It clearly follows that the one-loop contribution to $1/g^2$
from the massless states is not duality invariant.
Thus, these duality anomalies
must be cancelled by adding
new terms to the effective action.
Apart from a universal Green--Schwarz counter term \cite{DFKZ,CAOV},
the duality anomaly can be cancelled
by adding to eq.(1)  terms which describe the threshold contribution
due to the massive string states.
The threshold contributions are given in
terms of automorphic functions of the target space duality group.
Specifically, as described in ref.
\cite{FKLZ}, for general (2,2) Calabi--Yau
compactifications there exist two types of automorphic functions:
the first one provides a duality
invariant completion of $K_0$, whereas the second one
is needed to cancel the duality anomaly coming from
$\log\det K_{ij}^{R}$.

In the following let us concentrate on the first kind of automorphic
function. Consider the following norm $||\Delta(T_i)||^2=
\Delta(T_i){\rm e}^{K_0(T_i,\bar T_i)}\bar\Delta(\bar T_i)$
where $\Delta(T_i)$ is a holomorphic
section of a line bundle over the moduli space.
For $||\Delta||^2$ to be duality invariant
$\Delta$
has to transform as $\Delta\rightarrow \lbrack g(T_i)\rbrack^{-1}\Delta$.
In order to give  at least a formal construction of $\Delta$, one uses
the fact the moduli space of (2,2)
Calabi--Yau compactifications is a special
K\"ahler manifold.
The special K\"ahler potential can be written as
\begin{equation}
K_0=-\log ( X^I\bar{\cal F}_I+ \bar X^I{\cal F}_I).
\end{equation}
Here the $X^I$ ($I=0,\dots ,n$) are holomorphic
functions (sections) of the moduli: $X^I=X^I(T_i)$; ${\cal F}(X)$
is a homogeneous holomorphic function of degree 2
in the $n+1$ variables $X^I$ and ${\cal F}_I=\partial{\cal F}/\partial
X^I$.
Since the function ${|\Delta|^2{\rm e}^K}$ must be  duality-invariant,
$\Delta(T^i)$ must be a holomorphic section
of holomorphic degree one.
Holomorphic sections of exactly this degree are $X^I$  and ${\cal F}_I$.
Therefore, the most natural ansatz for $\Delta$ is to take a product
of all possible linear
combinations of these two holomorphic functions \cite{FKLZ}:
\begin{equation}
\log||\Delta ||^2=
 -\sum_{M_I,N^I}\log
{|M_IX^I+iN^I{\cal F}_I|^2
\over X^I\bar{\cal F}_I+\bar X^I{\cal F}_I}.
\end{equation}
(This sum has to be regularized in a suitable manner.)
Target space duality transformations act (up to a holomorphic
function) as symplectic
transformations on the vector $(X^I,i{\cal F}_I)$: $\Gamma\subset
Sp(2n+2,{\bf Z})$.
It follows that eq.(3)
provides a duality invariant
expression only if
the symplectic transformations on $X^I$ and $i{\cal F}_I$
are accompanied by the conjugate
symplectic  transformations on the integers
$M_I$ and $N^I$.
This means that $(M_I,N^I)$
must build proper representations of the symplectic
duality transformations, i.e. they form a restricted set of integers.

The only class of models
for which it is so far known how to solve explicitly
the constraint on the summation integers and how to regularize
the sum eq.(3)
are the orbifold compactifications \cite{DHVW}.
Let us consider,
for example, the simplest case with a single overall modulus
field
$T=-iX^1/X^0$. ($T=R^2+iB$ where $R$ is the overall
radius of the six-space and $B$ an internal axion.)
The corresponding
holomorphic function ${\cal F}(X)$ looks like ${\cal F}
=i(X^1)^3/ X^0$ implying, with eq.(2), $K_0=-3\log(T+
\bar T)$. Then the holomorphic section $\Delta$ takes the form:
\begin{equation}
\Delta(T)=
 \prod_{M_I,N^I}(M_0-iN^0T^3+iM_1T
+3N^1T^2)^{-1} .
\end{equation}
The action of the target space modular group
$\Gamma=PSL(2,{\bf Z})$ implies the following
restriction on the integers $M_I$ and $N^I$:
\begin{eqnarray}
M_0&=&m^3,\qquad N^0=n^3,\nonumber\\
M_1&=&3m^2n,\qquad N^1=-mn^2,
\end{eqnarray}
where $m$, $n$ are now unrestricted integers. Using $\zeta$-function
regularization (see ref. \cite{FKLZ}\ for details) $\Delta(T)$ finally
becomes
\begin{equation}
\Delta(T)=[ \prod_{m,n}(m+inT)^{-3} ]_{\rm reg}={1\over\eta(T)^6},
\end{equation}
where $\eta(T)={\rm e}^{-\pi T/12}\prod_{n=1}^\infty
(1-{\rm e}^{-2\pi nT})$
is the Dedekind function.

So let us discuss
the threshold effects in the framework of
orbifold compactifications
\cite{FILQ,DKL,DFKZ,LOUIS,ILR}.
The contribution, eq.(1), from the massless charged fields
plus the threshold effects, eq.(6),
of the massive orbifold excitations
leads to the following one-loop running gauge coupling constants
(up to the gauge group independent
Green--Schwarz term, and also up to a small $T$-independent term):
\begin{eqnarray}
{1\over g_a^2(\mu)}&=&{k_a\over  g_{\rm st}^2}+{b_a\over 16\pi^2}
\log{M_{\rm st}^2\over\mu^2}+
{1\over 16\pi^2}b_a'\log( T_R|\eta(T)|^4),\nonumber\\
b_a'&=&-b_a-2\sum_{R_a}h_{R_a}
T(R_a)(1+n_{R_a}).
\end{eqnarray}
where $T_R=T+\bar T=2R^2$, and
$b_a=-3C(G_a)+\sum_{R_a}h_{R_a}T(R_a)$ are the $N=1$ $\beta$-function
coefficients.
The integers $n_{R_a}$ are the modular weights of the massless
matter fields $\phi_i^{R_a}$, i.e. the matter fields
transform under $PSL(2,{\bf Z})$ as
$\phi_i^{R_a}\rightarrow\phi_i^{R_a}(icT+d)^{n_{R_a}}$.
For symmetric orbifold compactifications the $n_{R_a}$ generically
satisfy $-3\leq n_{R_a}\leq -1$.
Finally, $M_{\rm st}$ is the field independent string mass scale
\cite {KAPLU,DFKZ}:
\begin{equation}
M_{\rm st}=0.7\times g_{\rm st}  \times 10^{18}~{\rm GeV}.
\end{equation}

Now we are ready to discuss the unification of the gauge
coupling constants.
The unification mass scale $M_X$ where two gauge
group coupling constants are equal, i.e. ${1\over k_ag_a^2(M_X)}=
{1\over k_bg_b^2(M_X)}$, becomes using eq.(7),
\begin{equation}
M_X= M_{\rm st}
\lbrack
T_R|\eta(T)|^4\rbrack^{{b_a'k_b-b_b'k_a
\over 2(b_ak_b-b_bk_a)}}.
\end{equation}
Since the function
$T_R|\eta(T)|^4$ is smaller than one for all $T$
it follows that $M_X/M_{\rm st}$ is smaller (bigger) than one if
${b_a'k_b-b_b'k_a\over b_ak_b-b_bk_a}$ is bigger (smaller) than zero.
For example, consider
the symmetric ${\bf Z}_2\times{\bf Z}_2$ orbifold.
Here we find
that $b_a'=-b_a$ ($a=E_8,E_6$) ($n_{27}=-1$).
Therefore the unification scale is given by
$M_X=M_{\rm st}/(T_R^{1/2}|\eta(T)|^2)$ and is
larger than the string scale for all values of the radius.

Now let us apply the above discussion
to the case of the
unification of the three physical coupling constants
$g_1,g_2,g_3$ \cite{ILR}.
Namely, we will consider a possible situation in which
(i) the massless
particles with standard model gauge couplings are just those of the
minimal supersymmetric Standard Model with gauge group
$G={\rm SU}(3)\times{\rm SU}(2)\times{\rm U}(1)$;
(ii) there is no partial
(field theoretical) unification scheme below the string scale.
This situation is what we call {\it minimal string unification}.
Until now no realistic string model with these
characteristics has been built, but the model search done up to now
has been extremely limited and by no means complete.
The relevant evolution
of the electro-weak and strong coupling constants in the field theory
was considered in ref. \cite{DG}.
The experimental results for $\sin^2\theta _{\rm W}\simeq 0.233$ and
$\alpha_{\rm s}\simeq 0.11$ are in very good agreement
with data for a value of the unification mass
$M_X \simeq 10^{16}~{\rm GeV}$.
So we want to discuss the question whether
the
unification scale of the minimal supersymmetric Standard Model
can be made consistent
with the relevant string unification scale $M_{\rm st}$ eq.(8)
($g_{\rm st}\simeq 0.7$).
At first sight, this seems very unlikely since $M_{\rm st}$
is substantially
larger than the minimal SUSY model scale $M_X$. Minimal string
unification without threshold effects would predict (taking the value
$(\alpha_e)^{-1}=128.5$ as an input)
$\sin^2\theta_{\rm W}^0=0.218$ and
$\alpha_{\rm s}^0=0.20$ in gross conflict with the experimental data.

However, we will now show
that for orbifold compactifications
under rather constrained circumstances
the effects of the string threshold contributions
could make
the separation of these two scales consistent.
We will make use of the threshold
formulae of eqs.(7) and (9),
although they were originally derived for
a general class of $(2,2)$ orbifolds.
However, these formulae seem also to be valid
in the presence of Wilson lines and for $(0,2)$ types of gauge
embeddings. These types of models may, in
general, yield strings with the gauge group of the Standard Model
and appropriate matter fields as discussed, for example, in ref.
\cite{IMNQ}.

Using eqs.(7) and (9) the equations for the scale dependent electroweak
angle and the strong coupling constant can be
written after some standard algebra as ($k_2=k_3=1$, $k_1=5/3$)
\cite{ILR}
\begin{eqnarray}
\sin^2\theta_{\rm W}(\mu)&=&{3\over 8} -
{5\alpha_e\over 32\pi} A \log\biggl({M_{\rm st}^2\over \mu ^2}\biggr)   -
{5\alpha_e\over 32\pi} ( \delta A-A) \log(T_R|\eta (T)|^4),\nonumber\\
{1\over \alpha_{\rm s}(\mu )}& =&{3\over 8\alpha_e} -
{3\over 32\pi}B \log\biggl({M_{\rm st}^2\over \mu ^2}\biggr)  -
{3\over 32\pi } ( \delta B-B) \log(T_R|\eta (T)|^4)
\end{eqnarray}
where
$A={3\over 5}b_1-b_2=28/5$ and $B=b_1+b_2-{8\over 3}b_3=20$.
Denoting by $n^i_{\beta }$ the modular weight of the $i$-th generation
field of type $\beta =Q,U,D,L,E$ one finds explicitly
\begin{eqnarray}
\delta A&=&{2\over 5}\sum_{i=1}^{N_{\rm g}} (
7n^i_{\rm Q}+n^i_{\rm L}-4n^i_{\rm U}-n^i_{\rm D}-3n^i_{\rm E})
+{2\over 5} (2+n_{\rm H}+n_{\bar{\rm H}}),\nonumber\\
\delta B& =&2\sum_{i=1}^{N_{\rm g}} (n^i_{\rm Q}+n^i_{\rm D}
- n^i_{\rm L}-n^i_{\rm E}
) -2 (2+n_{\rm H}+n_{\bar{\rm H}})
\end{eqnarray}
where $N_{\rm g}$ is the number of generations and
$n_{\rm H}$ and $n_{\bar{\rm H}}$
are the modular weights of the Higgs fields.
Now we can search for the modular weights $ n_{\beta }$
leading to the correct experimental values for
$\alpha_{\rm s}(M_{\rm Z})$
and $\sin^2\theta_{\rm W}(M_{\rm Z})$ within their experimental
errors.
Assuming generation independence for the $n_{\beta }$ as well as
$-3\leq n_{\beta } \leq -1 $  one finds, interestingly enough,
a unique answer for the matter fields:
\begin{equation}
n_{\rm Q}=n_{\rm D}=-1\ \ ,\ \ n_{\rm U}=-2 \ \ ,\ \  n_{\rm L}
=n_{\rm E}=-3,
\end{equation}
and a constraint $n_{\rm H}+n_{\bar{\rm H}}=-5,-4$.
For $n_{\rm H}+n_{\bar{\rm H}}=-5$
one obtains
$\delta A=42/5$, $\delta B=30$,
and the three coupling constants meet at a scale $M_X\sim 2\times
10^{16}{\rm GeV}$ provided
that ${\rm Re}T\sim 16$.
For $n_H+n_{\bar H}=-4$ one has
$\delta A=44/5$, $\delta B=28$. Now the three couplings only meet
approximately for similar values of ${\rm Re}T$.
Thus we have
just shown that the {\it minimal string unification} scenario
is in principle compatible with the measured low energy coupling
constants
for, (i) sufficiently large ${\rm Re}T$, and
(ii) restricted choices of the
modular weights of the Standard Model particles.

If no string model with the characteristics of the above
minimal unification scenario is found, one can think about the
following two alternatives with possibly only tiny
string threshold effects:
(i) There is
unification at the
scale $M_X\sim 10^{16}$ GeV, but with an intermediate field theory
GUT group between $M_X$ and $M_{\rm st}$. (ii) There
is  string unification at $M_X\simeq M_{\rm st}$ with
extra light states between the weak scale and $M_{\rm st}$
contributing to the field theory running of the coupling constants
\cite{PLANCK}. A combination of these two alternatives
is of course also possible as investigated in the context of the
flipped ${\rm SU}(5)$ model \cite{FLIPP}.
In any case, it is clear that
the present precision of the measurement of low energy
gauge couplings has reached a level which is sufficient to
test some fine details of string models.

\vskip0.5cm
I would like to thank S. Ferrara, L. Ib\'a\~nez, C. Kounnas,
G. Ross and F. Zwirner for their very fruitful collaboration
on the topics presented in this paper.

\def\refbreak{\hfil\penalty200\hfilneg}
\def\nup#1({\refbreak\ Nucl.\ Phys.\ ${\underline{B#1}}$\ (}
\def\plt#1({\refbreak\ Phys.\ Lett.\ ${\underline{B#1}}$\ (}
\def\plb#1({\refbreak\ Phys.\ Lett.\ ${\underline{#1B}}$\ (}
\def\rB{\hfil\penalty1000\hfilneg}


\begin{thebibliography}{99}
\bibitem{KIK} K. Kikkawa and M. Yamasaki, \plb149 (1984) 357;
               N. Sakai and I. Senda, Progr. Theor. Phys. 75 (1986) 692.

\bibitem{FLST} S. Ferrara, D. L\"ust, A. Shapere and S. Theisen,
           \plt225 (1989) 363; S. Ferrara, D. L\"ust and S. Theisen,
           \plt233 (1989) 147.

\bibitem{KAPLU} V. Kaplunovsky, \nup307 (1988) 145.

\bibitem{CHSW} P. Candelas, G. Horowitz, A. Strominger and E. Witten,
       \nup258 (1985) 46.

\bibitem{DFKZ} J.P. Derendinger, S. Ferrara, C. Kounnas
       and F. Zwirner, preprint CERN-TH.6004/91 (1991).

\bibitem{LOUIS} J. Louis, preprint  SLAC-PUB-5527 (1991).

\bibitem{CAOV} G. Cardoso and B. Ovrut, preprint UPR-0464T (1991).

\bibitem{FKLZ} S. Ferrara, C. Kounnas, D. L\"ust and F. Zwirner,
                               preprint CERN-TH.6090/91 (1991),
           to appear in Nucl. Phys. B.

\bibitem{DHVW} L. Dixon, J. Harvey, C.~Vafa and E.~Witten,
         \nup261 (1985) 651; \nup274 (1986) 285.

\bibitem{FILQ} A. Font, L.E. Ib\'a\~nez, D. L\"ust and F. Quevedo,
           \plt245 (1990) 401;
           S. Ferrara, N. Magnoli, T.R. Taylor and
           G. Veneziano, \plt245 (1990) 409.

\bibitem{DKL} L. Dixon, V. Kaplunovsky and J. Louis,
         \nup355 (1991) 649;
                  I. Antoniadis, K.S. Narain and T.R. Taylor,
         \plt267 (1991) 37.

\bibitem{ILR} L. Ib\'a\~nez, D. L\"ust and G. Ross, preprint
               CERN-TH.6241/91 (1991), to appear in Phys. Lett. B.

\bibitem{DG} S. Dimopoulos, S. Raby and F. Wilczek, Phys. Rev.
D24 (1981) 1681; L.E. Ib\'a\~nez and G.G. Ross, \plb105 (1981) 439;
S. Dimopoulos and H. Georgi, \nup193 (1981) 375;
J. Ellis, S. Kelley and D.V. Nanopoulos, \plt249 (1990) 441;
\plt260 (1991) 131;
P. Langacker, preprint UPR-0435T, (1990);
U. Amaldi, W. de Boer and H. F\"urstenau, \plt260 (1991) 447;
P. Langacker and M. Luo, preprint UPR-0466T, (1991).

\bibitem{IMNQ} L.E. Ib\'a\~nez, H.P. Nilles and F. Quevedo,
\plt187 (1987) 25; L.E. Ib\'a\~nez, J. Mas, H.P. Nilles and
F. Quevedo, \nup301 (1988) 157; A. Font, L.E. Ib\'a\~nez,
F. Quevedo and A. Sierra, \nup331 (1990) 421.

\bibitem{PLANCK} L.E. Ib\'a\~nez, \plb126 (1983) 196;
J.E. Bjorkman and D.R.T. Jones, \nup259 (1985) 533;
I. Antoniadis, J. Ellis, S. Kelley and D.V. Nanopoulos,
preprint CERN-TH.6169/91 (1991).

\bibitem{FLIPP} I. Antoniadis, J. Ellis, R. Lacaze
and D.V. Nanopoulos, preprint CERN-TH.6136/91 (1991);
    S. Kalara, J.L. Lopez and D.V. Nanopoulos,
      preprint CERN-TH-6168/91 (1991).

\end{thebibliography}
\end{document}